\documentclass[12pt,a4paper,oneside,draft]{amsart}
\usepackage{amsmath, amsfonts,amssymb}

\newtheorem{theorem}{Theorem}[section]
\newtheorem{lemma}[theorem]{Lemma}
\newtheorem{proposition}[theorem]{Proposition}
\newtheorem{corollary}[theorem]{Corollary}

\theoremstyle{definition}
\newtheorem{definition}[theorem]{Definition}
\newtheorem{example}[theorem]{Example}

\theoremstyle{remark}
\newtheorem{remark}[theorem]{Remark}

\numberwithin{equation}{section}

\newcommand{\cD}{\mathcal D}
\newcommand{\cF}{\mathcal F}
\newcommand{\cL}{\mathcal L}
\newcommand{\cS}{\mathcal S}

\newcommand{\A}{\mathbb{A}}
\newcommand{\C}{\mathbb{C}}
\newcommand{\LC}{\mathbb{L}}
\newcommand{\M}{\mathbb{M}}
\newcommand{\N}{\mathbb{N}}
\newcommand{\R}{\mathbb{R}}
\newcommand{\Z}{\mathbb{Z}}

\newcommand{\Mbil}[2]{\left<#1,#2\right>_\M}
\newcommand{\LCbil}[2]{\left<#1,#2\right>_\LC}

\newcommand{\bx}{\mathbf{x}}
\newcommand{\by}{\mathbf{y}}
\newcommand{\bp}{\mathbf{p}}

\newcommand{\LCBox}{\widetilde{\Box}}
\newcommand{\LCphi}{\widetilde{\phi}}
\newcommand{\LCOmega}{\widetilde{\Omega}}
\newcommand{\LComega}{\tilde{\omega}}
\newcommand{\LCGamma}{\widetilde{\Gamma}}
\newcommand{\LCpi}{\widetilde{\pi}}

\newcommand{\LCcS}{\cS_{\partial_-}}
\newcommand{\LCb}{\tilde{b}}
\newcommand{\LCp}{\tilde{p}}
\newcommand{\LCu}{\tilde{u}}
\newcommand{\LCv}{\tilde{v}}
\newcommand{\LCx}{\tilde{x}}
\newcommand{\LCy}{\tilde{y}}
\newcommand{\LCbp}{{\tilde{\bp}}}
\newcommand{\LCbx}{{\tilde{\bx}}}
\newcommand{\LCby}{{\tilde{\by}}}
\newcommand{\LCD}{{\widetilde{D}}}
\newcommand{\LCQ}{\widetilde{Q}}
\newcommand{\LCS}{\tilde{S}}
\newcommand{\isomto}{\stackrel{\sim}{\to}}
\newcommand{\hatM}{{\wedge \M}}
\newcommand{\hatL}{{\wedge \LC}}
\newcommand{\pLCFT}{\cF_\LC^{\LCbx\to\LCbp}}
\newcommand{\LCRpm}[1]{\R_{\gtrless 0}\times \R^{#1}}
\newcommand{\LChat}[1]{{#1}^\sqcap}

\newcommand{\supp}{\mbox{supp}}
\newcommand{\injto}{\hookrightarrow}
\newcommand{\PV}{\mbox{PV}}
\newcommand{\tres}{|^*}

\begin{document}
\title{Uniqueness in the Characteristic Cauchy
Problem of the Klein-Gordon
Equation and Tame Restrictions of Generalized Functions}
\author{Peter Ullrich}
\address{Institut f\"ur Informatik, TU M\"unchen,
Boltzmannstr. 3, D-85748 Garching, Germany \and
Institut f\"ur Physik, Universit\"at Regensburg, Germany}
\email{ullrichp@in.tum.de}
\begin{abstract}
We show that every tempered distribution, which is a solution of the
(homogenous) Klein-Gordon equation, admits a ``tame'' restriction to the
characteristic (hyper)surface
$\{x^0+x^n=0\}$ in $(1+n)$-dimensional Minkowski space and
is uniquely determined by this restriction. The
restriction belongs to the space $\cS'_{\partial_-}(\R^n)$
which we have introduced in \cite{PullJMP}. Moreover, we show that every
element of $\cS'_{\partial_-}(\R^n)$ appears as the ``tame'' restriction of a
solution of the (homogeneous) Klein-Gordon equation.
\end{abstract}
\maketitle
%
\section{Introduction}
%
\noindent
The characteristic Cauchy problem of the Klein-Gordon equation asks for
solutions $u$ of the (homogeneous) Klein-Gordon equation $(\Box+m^2)u=0$,
where $\Box=\partial^2_0-\sum_{i=1}^n\partial^2_i$,  with
prescribed initial data on the surface $\Sigma=\{x^0+x^n=0\}$. Since $\Sigma$ is a 
characteristic of the Klein-Gordon operator, the general
theory of (linear) partial differential equations predicts 
non-uniqueness of the solutions unless growth conditions are imposed
\cite{Hoer2, Hoer1}. 

The study of the characteristic Cauchy problem is motivated by light cone
quantum field theory. In contrast to classical quantum field theory which takes place
in Minkowski space-time with coordinates $x=(x^0,x^1,x^2,x^3)$, where
$x^0$ takes on the role of time, light cone quantum field theory uses a different
coordinate system obtained by a linear change of variables,
where $x^+=(1/ \sqrt 2)(x^0+x^3)$ is the new time variable
The use of this new set of variables, especially the use of
$x^+$ as time (evolution) parameter, results in the use of a new kind of dynamics
-- called {\em front form dynamics} -- which was introduced
 by P.A.M. Dirac in \cite{Dirac}. The use of this different kind of dynamics is just the
starting point of light cone quantum field theory \cite{BrosPins}.
Hence, one naturally arrives at the  characteristic Cauchy problem when considering
fields in the framework of light cone field theory. The problem that the initial data is given
on a characteristic surface was widely seen as a big disadvantage of light cone
field theory since its beginning. However, Leutwyler et al.\ \cite{LKS} 
assumed that at least within the space of physical solutions uniqueness holds true without
giving a proof. Recently, Heinzl and Werner \cite{HeinzlWerner} have shown a 
uniqueness result in the special situation of 1+1 dimensions considering solutions
enclosed in a box with various kinds of boundary conditions.

In \cite{PullJMP} we have introduced a novel topological vector space $\LCcS(\R^n)$ 
along with its dual space $\LCcS'(\R^n)$ in connection with the fundamental problem
of light cone quantum field theory that the real scalar free field admits no canonical
restriction to $\{x^0+x^3=0\}$.
 In this paper we will use the space $\LCcS'(\R^n)$ to define for
each solution $u\in\cS'(\R^{1+n})$ of the Klein-Gordon equation $(\Box+m^2)u=0$
a non-canonical, ``tame'' restriction $u_0\in\LCcS'(\R^n)$ to the
hypersurface $\Sigma$ in Minkowski space, and show that $u$ is
uniquely determined by $u_0$. Moreover, we show that each $u_0\in\LCcS'(\R^n)$
appears as the ``tame'' restriction of some $u\in\cS'(\R^{1+n})$ solving the
Klein-Gordon equation.
%
\section{Notation and conventions}
%
Let $\M=\M^{1+n}$ denote
$(1+n)$-dimensional Minkowski space, i.e., Euclidean space $\R^{1+n}$
together with the bilinear form $\Mbil{x}{y}=x^0y^0-\sum_{i=1}^nx^i
y^i$. We distinguish the variable $x^0$ and write $x=(x^0,\bx)\in\M$, where
$\bx=(x^1,\ldots,x^n)$. For $\bx,\by\in\R^n$ we denote
by $\bx\cdot\by=\sum_{i=1}^nx^iy^i$ their Euclidean scalar product, hence
$\Mbil{x}{y}=x^0y^0-\bx\cdot\by$. Furthermore, we set $x^2=\Mbil{x}{x}$,
$\bx^2=\bx\cdot\bx$ and $|\bx|=\sqrt{\bx^2}$. If $f$ is an integrable
(complex-valued) function on Minkowski space $\M$, we denote by
$\cF_\M f=f^\hatM$ the Fourier transform of $f$ with respect to the Minkowski
bilinear form, i.e.,$(\cF_\M f)(p)=f^\hatM(p)=\int
dx f(x)e^{i\Mbil{x}{p}}$, whereas
$\cF f=f^\wedge$ denotes the Fourier transform of $f$ with
respect to Euclidean scalar product, i.e., $(\cF f)(\bp)=f^\wedge(\bp)
=\int d\bx f(\bx)e^{-i\bx\cdot\bp}$. Recall, that by the inversion formula
$f^{\wedge\wedge}=(2\pi)^n f^\lor$, where $f^\lor(x)=f(-x)$.

In the characteristic Cauchy problem of the Klein-Gordon equation
the initial data is given on the characteristic surface 
$\Sigma=\{x^0+x^n=0\}\subset\M$. It is appropriate to go over to
light-cone coordinates $\LCx=(\LCx^0,\ldots,\LCx^n)$
according to the linear transformation $\LCx=\kappa(x)$ given by
$\LCx^0=(1/\sqrt 2)(x^0+x^n)$, $\LCx^i=x^i$ ($i=1,\ldots,n-1$),
$\LCx^n=(1/\sqrt 2)(x^0-x^n)$.
Usually, in physical literature, the components of
$\LCx$ are denoted by $\LCx=(x^+,x_\bot,x^-)$, where $x^+=\LCx^0$,
$x_\bot=(\LCx^1,\ldots,\LCx^{n-1})$ and $x^-=\LCx^n$. We will use mainly this
notation from physics. We also distinguish the variable $x^+$ --
the LC-time-variable -- and write $\LCx=(x^+,\LCbx)$, where $\LCbx=(x_\bot,x^-)$.
The transformation $\kappa$ maps $\Sigma$ onto $\{x^+=0\}$. Furthermore,
the Minkowski bilinear form is transformed to the LC-bilinear form
$\LCbil{\LCx}{\LCy}=x^+y^-+x^-y^+ - x_\bot\cdot y_\bot$ by $\kappa$, where
$x_\bot\cdot y_\bot=\sum_{i=1}^{n-1}\LCx^i\LCy^i$. We set 
$\LCx^2=\LCbil{\LCx}{\LCx}=2x^+x^- - x_\bot\cdot x_\bot$.
We denote $\LC=\LC^{1+n}$
the bilinear space consisting of $\R^{1+n}$ and $\LCbil{.}{.}$, and call
it $(1+n)$-dimensional {\em LC-space}. Hence, $\kappa:\M\isomto\LC$ is an
isomorphism of bilinear spaces. If $f$ is an integrable (complex-valued)
function on LC-space $\LC$, we denote by $\cF_\LC f=f^\hatL$
the Fourier transform of $f$ with respect to the LC-bilinear form, i.e.,
$(\cF_\LC f)(\LCp)=f^\hatL(\LCp)=\int d\LCx f(\LCx)e^{i\LCbil{\LCx}{\LCp}}$.
Notice that $\kappa$ commutes with these Fourier
transformations,
i.e., $(f\circ\kappa)^\hatM=f^\hatL\circ\kappa$. Next we need to
introduce a further Fourier transformation which affects only the spatial part
of $\LCx=(x^+,\LCbx)$. 
Since in the LC-bilinear form
$\LCbil{x}{p}=x^+p^-+x^-p^+-x_\bot\cdot p_\bot$ the time-variable $x^+$
is paired with $p^-$, the variable $p^-$ will be considered
as energy-variable in physical literature. Hence the variable
$p=(p^+,p_\bot,p^-)$ is split into $p=(\LCbp,p^-)$, where
$\LCbp=(p^+,p_\bot)$ denotes the spatial momentum. In contrast, we
split $x$ into $x=(x^+,\LCbx)$, since we have distinguished $x^+$ as
time-variable. Here, a little bit care is needed. If $f:\R^n\to\C$ is
an (absolutely) integrable function, we set
\begin{equation}\label{equ:partial_LC-Fourier}
    (\pLCFT f)(\LCbp)=\LChat f(\LCbp)=
        \int d^n\LCbx f(\LCbx)e^{i(x^-p^+-x_\bot\cdot p_\bot)}.
\end{equation}
Let $\cS(\R^n)$ denote the Schwartz space consisting of rapidly decreasing,
smooth, complex-valued functions on $\R^n$, i.e., complex-valued
$C^\infty$-functions $f$ on $\R^n$ such that
$\sup(1+|\bx|)^N|\partial^\alpha f(\bx)|<\infty$
for all $N\in\N$ and multi-indices $\alpha$. $\cS(\R^n)$ is topologized by
the family of seminorms $\sup(1+|\bx|)^N|\partial^\alpha .|$ ($N\in\N,
\alpha$ multi-index). The dual space $\cS'(\R^n)$ is called the space
of generalized functions (or tempered distributions).
Usually, $\cS'(\R^n)$ carries the weak$^*$-topology. There is also a canonical
embedding $\cS(\R^n)\injto\cS'(\R^n)$ and, in the sequel, we often identify
$\cS(\R^n)$ with its image, i.e., we assume $\cS(\R^n)\subset\cS'(\R^n)$.
As is well known (cf. e.g. \cite{Rudin_func}),
the Fourier
transformation $\cF$ is a linear homeomorphism
from $\cS(\R^n)$ onto $\cS(\R^n)$ which extends to a linear, sequentially
continuous mapping from $\cS'(\R^n)$ onto $\cS'(\R^n)$. Obviously, the
same holds for $\cF_\M$, $\cF_\LC$ and $\pLCFT$. We denote by $\cD(U)$
$(U\subset\R^n)$ the topological vector space of all complex-valued
smooth, i.e., $C^\infty$ functions
on $U$ with compact support, and by $\cD'(U)$ its dual space -- the space
of distributions \cite{Hoer1, Rudin_func}.

Furthermore, we set $\Gamma^\pm_m=\{p\in\M^{1+n}:p^2-m^2=0,~\pm p^0>0\}$,
$\LCGamma^\pm_m=\{\LCp\in\LC^{1+n}:\LCp^2-m^2=0,~\pm p^->0\}$ and
$\Omega_\pm:\R^n\to\Gamma^\pm_m$, $\bp\mapsto(\pm\omega(\bp), \bp)$,
$\omega(\bp)=\sqrt{\bp^2+m^2}$ and
$\LCOmega_\pm:\LCRpm{n-1}\to\LCGamma^\pm_m$,
$\LCbp\mapsto(\LCbp,\LComega(\LCbp))$, $\LComega(\LCbp)=(\LCbp,\LComega(\LCbp))$,
$\LComega(\LCbp)=(1/2p^+)(p_\bot^2+m^2)$, and
$\LCOmega(\LCbp)=(\LCbp,\LComega(\LCbp))$ ($\LCbp\in\R^n\setminus\{p^+=0\}$).
%
\section{Review of the non-characteristic Cauchy problem}
\label{sec:rev_Cauchy}
%
It is well known (see, e.g., \cite{Bog1}) that the following 
(non-characteristic) Cauchy problem of the Klein-Gordon equation,
stated in $\cS'(\R^{1+n})$, is well-posed. Let $\Box=\partial_0^2-\Delta_\bx$
denote the d'Alembert operator. Then a solution $u\in\cS'(\R^{1+n})$ of
\begin{equation}\label{equ:non-char_Cauchy}
    (\Box+m^2)u=0,\qquad u|_{x^0=0}=u_0,\qquad\partial_0u|_{x^0=0}=u_1.
\end{equation}
exists and is unique for any $u_0,u_1\in\cS'(\R^n)$. Since in the following
sections we essentially make use of a special representation of the solutions
of \eqref{equ:non-char_Cauchy}, we will give a short review of the
non-characteristic Cauchy problem \eqref{equ:non-char_Cauchy}.
First of all, we
have to note that any solution $u\in\cS'(\R^{1+n})$ of the Klein-Gordon
equation is $C^\infty$-dependent on $x^0\in\R$ as a parameter
\cite{Bog1}, and hence admits a restriction $u|_{x^0=0}$ to
$\{x^0=0\}$. This follows easily from the fact that the Klein-Gordon
operator is hypoelliptic with respect to $x^0$
\cite{Garding1958}, \cite{Garding1961}, \cite{Ehrenpreis1962}. Thus one obtains
a uniquely determined family $(u_{x^0})_{x^0\in\R}$ in $\cS'(\R^n)$ such that
\begin{equation}\label{equ:parameter}
    (u(x),f(x^0)g(\bx))=\int (u_{x^0},g)f(x^0)dx^0
\end{equation}
for all $f(x^0)\in\cS(\R)$, $g(\bx)\in\cS(\R^n)$. The restriction $u|_{x^0=0}$
is then, per definition, $u_{x^0=0}$. Usually, the uniqueness of a solution
of \eqref{equ:non-char_Cauchy} is proven by showing that the 
(parameter) derivatives of any order of $u_{x^0}$ with respect to $x^0$
vanish at $x^0=0$, and that the family $(u_{x^0})_{x^0\in\R}$
depend analytically
on $x^0$ (the last assertion follows by a theorem of Paley and Wiener
\cite{Rudin_func}). These arguments are not directly applicable to
the characteristic Cauchy problem. Hence we will reprove uniqueness and
existence of \eqref{equ:non-char_Cauchy} in a manner which is more in
the spirit of the proof of uniqueness and existence of the characteristic Cauchy problem.
Moreover, we need the following results as a preparation for studying
the connection between the characteristic and the non-characteristic Cauchy problem
in Subsection~\ref{sub:relation}.
\begin{definition}
For each $a(\bp)\in\cS'(\R^n)$ we define the generalized functions
$a(\bp)\delta_\pm(p^2-m^2)\in\cS'(\R^{1+n})$ by
$$
    (a(\bp)\delta_\pm(p^2-m^2),f(p))=\left(a(\bp),
    \frac{f(\Omega_\pm(\bp))}{2\omega(\bp)}\right),
    \qquad (f(p)\in\cS(\R^{1+n})),
$$
where $\Omega_\pm(\bp)=(\pm\omega(\bp),\bp)$ and
$\omega(\bp)=\sqrt{\bp^2+m^2}$ ($\bp\in\R^n$).
\end{definition}
\begin{remark}\label{rem:r1}
a) Let $\Gamma_m=\{p\in\R^{1+n}:p^2-m^2=0\}$ be the
mass-hyperboloid. Then the projection map $\Gamma_m\to\R^n$,
$p=(p^0,\bp)\mapsto\bp$ is a double covering of $\R^n$. The restriction
of the projection map to each of the connected components
$\Gamma^\pm_m=\Gamma_m\cap\{\pm p^0>0\}$ of $\Gamma_m$ is a
homeomorphism $\Gamma^\pm_m\isomto\R^n$ whose inverse mapping is
$\R^n\to\Gamma^\pm_m$, $\bp\mapsto\Omega_\pm(\bp)$.

b) Since $f(p)\mapsto f(\Omega_\pm(\bp))/2\omega(\bp)$ are continuous, linear
maps from $\cS(\R^{1+n})$ to $\cS(\R^n)$, $a(\bp)\delta_\pm(p^2-m^2)$ are
well-defined tempered distributions.

c) We denote by $dS$ the canonical surface measure on $\Gamma_m$. If
$a(\bp)\in\cS(\R^n)\subset\cS'(\R^n)$ then
$$
    (a(\bp)\delta_\pm(p^2-m^2),f(p))=\int_{\Gamma^\pm_m}
    \frac{(a\circ\Omega^{-1}_\pm)(p)f(p)}{|\nabla Q(p)|}dS(p),
    \qquad (f(p)\in\cS(\R^{1+n})),
$$
where $Q(p)=p^2-m^2$.

d) The support of 
$a(\bp)\delta_\pm(p^2-m^2)$ is contained in $\Gamma^\pm_m$.
\end{remark}
The following lemma is an immediate consequence of the definition.
\begin{lemma}
The mappings $\cS'(\R^n)\to\cS'(\R^{1+n})$, 
$a(\bp)\mapsto a(\bp)\delta_\pm(p^2-m^2)$ are $\C$-linear and sequentially
continuous.
\end{lemma}
\begin{lemma}\label{lem:inj}
Let $a_0(\bp),a_1(\bp),a(\bp)\in\cS'(\R^n)$.

(i) $a(\bp)\delta_\pm(p^2-m^2)=0$ if and only if $a(\bp)=0$.

(ii) $a_0(\bp)\delta_+(p^2-m^2)+a_1(\bp)\delta_-(p^2-m^2)=0$
if and only if $a_0(\bp)=a_1(\bp)=0$.
\end{lemma}
\begin{proof}
(i) This follows from the fact, that the maps 
$$	\cS(\R^{1+n})\to\cS(\R^n),~
	f(p)\mapsto f(\Omega_\pm(\bp))/2\omega(\bp)
$$
are onto. For, given $g(\bp)\in\cS(\R^n)$ then
$f(p)=2\omega(\bp)e^{-(p^0\mp\omega(\bp))^2}g(\bp)\in\cS(\R^{1+n})$ does the
job.

(ii) From $a_0(\bp)\delta_+(p^2-m^2)+a_1(\bp)\delta_-(p^2-m^2)=0$ it
follows $a_0(\bp)\delta_+(p^2-m^2)=-a_1(\bp)\delta_-(p^2-m^2)$. Now,
since the supports are disjoint (cf. Remark~\ref{rem:r1}\,(d)),
the assertion follows from (i).
\end{proof}
\begin{proposition}\label{prop:general_solution1}
The general solution of the division problem $(p^2-m^2)u=0$, 
$u\in\cS'(\R^{1+n})$ is given by
$$
    u(p)=a_0(\bp)\delta_+(p^2-m^2) + a_1(\bp)\delta_-(p^2-m^2), 
$$
where $a_0(\bp),a_1(\bp)\in\cS'(\R^n)$. Moreover, $a_0(\bp)$ and $a_1(\bp)$
are uniquely determined by $u(p)$.
\end{proposition}
\begin{proof}
Uniqueness follows from Lemma~\ref{lem:inj}. To prove existence
(see also e.g. \cite{Bog1}, p. 60) let $u\in\cS'(\R^{1+n})$ be a solution
of the division problem. Since $\supp(u)\subset\Gamma_m$ and $\Gamma_m$
has the two connected components $\Gamma^\pm_m=\{p\in\Gamma_m:\pm p^0>0\}$
we can uniquely split $u=u_+ + u_-$ with $u_\pm=u|_{\pm p^0>0}$. Hence
in the following we may assume w.l.o.g.\ that $u=u_+$, i.e.,
$\supp(u)\subset\Gamma^+_m$. Consider the smooth coordinate transformation
$$
	\lambda:\R_{>0}\times\R^n, (p^0,\bp)\mapsto(t,\bp)=(p^2-m^2,\bp)
$$
which maps $\Gamma^+_m$ onto $\{0\}\times\R^n$. Since $u$ has support
in $\Gamma^+_m$ $\lambda_*u=u\circ\lambda^{-1}$ has support in
$\{0\}\times\R^n$. Hence we can write $\lambda_*u=\delta(t)\otimes a_0(\bp)$
with some $a_0(\bp)\in\cS'(\R^n)$, and thus
$u=\lambda^*(\delta(t)\otimes a_0(\bp))$. By the general formula
for smooth coordinate changes (see, e.g., \cite{Rudin_func, Hoer1}) we obtain
\begin{align*}
	(\lambda^*(\delta(t)\otimes a_0(\bp)),f) & =
	(\delta(t)\otimes a_0(\bp),
	  \frac{f(\sqrt{t+\bp^2+m^2},\bp)}{2\sqrt{t+\bp^2+m^2}})\\
	  & = (a_0(\bp),\frac{f(\omega(\bp),\bp)}{2\omega(\bp)})\\
	  & = (a_0(\bp)\delta_+(p^2-m^2),f)
\end{align*}
for all $f\in\cS(\R^{1+n})$.
\end{proof}
\begin{proposition}\label{prop:main_non-char_Cauchy}
Assume $a_0(\bp),a_1(\bp)\in\cS'(\R^n)$ and let 
$u(x)=\cF^{-1}_\M(a_0(\bp)\delta_+(p^2-m^2)+a_1(\bp)\delta_-(p^2-m^2))\in
\cS'(\R^{1+n})$. Then $u(x)=u(x^0,\bx)$ is $C^\infty$-dependent on
$x^0$ as a parameter, and for all $x^0\in\R$
\begin{align*}
    u_{x^0} & =\frac{1}{4\pi}\cF^{-1} \left(
        \frac{a_0(\bp)e^{-i\omega(\bp)x^0} + a_1(\bp)e^{i\omega(\bp)x^0}}
        {\omega(\bp)}
        \right),\\
    (\partial_0 u)_{x^0} & =\frac{1}{4\pi i}\cF^{-1} \left(
        a_0(\bp)e^{-i\omega(\bp)x^0} - a_1(\bp)e^{i\omega(\bp)x^0}
        \right).
\end{align*}
\end{proposition}
\begin{proof}
We define the family $(u_{x^0})_{x^0\in\R}$ by
$u_{x^0}=\frac{1}{4\pi}\cF^{-1}\left(
\frac{a_0(\bp)e^{-i\omega(\bp)x^0} + a_1(\bp)e^{i\omega(\bp)x^0}}
{\omega(\bp)}\right)\in\cS'(\R^n)$. Let $f(x^0)\in\cS(\R)$ and
$g(\bx)\in\cS(\R^n)$. We have to show that $u(\bx)$ fulfills the
equation~\eqref{equ:parameter}.

\textbf{Case 1:} Firstly, we assume $a_0(\bp),a_1(\bp)\in\cS(\R^n)\subset
\cS'(\R^n)$. Then, by an easy computation, we obtain
\begin{align*}
  (u, f\otimes g) & = \left(
   a_0(\bp)\delta_+(p^2-m^2)+a_1(\bp)\delta_-(p^2-m^2),\cF^{-1}_\M(f\otimes g)
   \right )\\
   & =\frac{1}{(2\pi)^{n+1}}\int dx^0 f(x^0)\int d^n\bx g(\bx)
     \int\frac{d^n\bp}{2\omega(\bp)}\left(
     a_0(\bp)e^{-i\omega(\bp)x^0} + a_1(\bp)e^{i\omega(\bp)x^0}
     \right)e^{i\bp\cdot\bx}\\
   & =\frac{1}{2\pi}\int dx^0 f(x^0)
     \int\frac{d^n\bp}{2\omega(\bp)}\left(
     a_0(\bp)e^{-i\omega(\bp)x^0} + a_1(\bp)e^{i\omega(\bp)x^0}
     \right)(\cF^{-1}g)(\bp)\\
   & = \int dx^0 f(x^0) (u_{x^0},g).
\end{align*}

\textbf{Case 2:} Now, assume $a_0(\bp),a_1(\bp)\in\cS'(\R^n)$. Choose
sequences $(a^{(m)}_0(\bp))$ and $(a^{(m)}_1(\bp))$ in $\cS(\R^n)$
such that $a^{(m)}_0(\bp)\to a_0(\bp)$, $a^{(m)}_1(\bp)\to a_1(\bp)$
in $\cS'(\R^n)$ and define 
$u^{(m)}=\cF^{-1}_\M(a^{m}_0(\bp)\delta_+(p^2-m^2)+
a^{m}_1(\bp)\delta_-(p^2-m^2))\in\cS'(\R^{1+n})$. Then
$u^{(m)}$ converges to $u$ in $\cS'(\R^{1+n})$ and, by construction,
$u^{(m)}_{x^0}$ converges to $u_{x^0}$ in $\cS'(\R^n)$ ($m\to\infty$). By the first case,
we have
\begin{equation}\label{equ:main1}
    (u,f\otimes g) = \lim_{m\to\infty}\int (u^{(m)}_{x^0},g) f(x^0)dx^0.
\end{equation}
Since
$|(u^{(m)}_{x^0},g)|\leq(4\pi)^{-1}\int\frac{d\bp}{\omega(\bp)}
(|(a^{(m)}_0(\bp)|+|a^{(m)}_1(\bp)|)|\cF^{-1}g(\bp)|=C<\infty$ 
where the constant $C$ doesn't depend on $x^0$,
the right-hand side of \eqref{equ:main1} equals $\int (u_{x^0},g)f(x^0)dx^0$
by dominant convergence.
\end{proof}
\begin{corollary}
Any solution $u\in\cS'(\R^{1+n})$ of the Klein-Gordon equation 
$(\Box+m^2)u=0$ is $C^\infty$-dependent on $x^0\in\R$ as a parameter.
\end{corollary}
\begin{corollary}
The (non-characteristic) Cauchy problem \eqref{equ:non-char_Cauchy} has
a unique solution for any $u_0,u_1\in\cS'(\R^n)$.
\end{corollary}
\begin{proof}
Assume $u_0(\bx),u_1(\bx)\in\cS'(\R^n)$. Define
$a_0(\bp),a_1(\bp)\in\cS'(\R^n)$ by
$$
    a_0(\bp)=2\pi\left(\omega(\bp)\hat u_0(\bp)+ i\hat u_1(\bp)\right)
    \quad\text{and}\quad
    a_1(\bp)=2\pi\left(\omega(\bp)\hat u_0(\bp) - i\hat u_1(\bp)\right).
$$
Then $u=\cF^{-1}_\M(a_0(\bp)\delta_+(p^2-m^2) + a_1(\bp)\delta_-(p^2-m^2))$
is a solution of \eqref{equ:non-char_Cauchy} by 
Proposition~\ref{prop:main_non-char_Cauchy}. Uniqueness follows immediately
from Propositions~\ref{prop:general_solution1}
and~\ref{prop:main_non-char_Cauchy}.
\end{proof}
%
\section{Squeezed generalized functions and tame restrictions}
%
\subsection{Definitions and elementary properties}
In this subsection we introduce squeezed generalized functions and 
show some general properties of this class of functions which will be
important in the sequel. Since we have already introduced
squeezed generalized functions in \cite{PullJMP}, we only give a summary
of results and omit most of the proofs.
\begin{definition}\label{def:squeezed_functions}
(a) Let $\cS_{p^+}(\R^n)$ be the set of all $f\in C^\infty(\R^n,\C)$
such that
\begin{equation}\label{eq:def1}
    ||f||_{k,\beta,\alpha}=\sup_{(p^+,p_\bot)\in\R^n\setminus\{p^=0\}}
    |(p^+)^kp_\bot^\beta\partial^\alpha f(p^+,p_\bot)|<\infty
\end{equation}
for all $k\in\Z$ and all multi-indices $\alpha, \beta$. We endow
$\cS_{p^+}(\R^n)$ with the locally convex topology defined by the seminorms
$||\_\_||_{k,\beta,\alpha}$ and call it the {\em squeezed Schwartz space}
(of squeezed rapidly decreasing functions). The dual space $\cS'_{p^+}(\R^n)$
is called the space of {\em squeezed generalized functions}
(or {\em squeezed tempered distributions}).

(b) We call $\cS_{p^+\gtrless 0}(\R^n)=\{f\in\cS_{p^+}(\R^n):
f|_{p^+\lessgtr 0}\equiv 0\}$ the {\em positive/negative squeezed
Schwartz space} (of positive/negative squeezed rapidly decreasing
functions). The dual space $\cS'_{p^+\gtrless 0}(\R^n)$ is called the space
of {\em positive/negative squeezed generalized functions}.

(c)
On $\cS'_{p^+}(\R^n)$ and $\cS'_{p^+\gtrless 0}(\R^n)$ we consider always
the $\text{weak}^*$-topology, i.e., the locally convex topology defined by the family
of seminorms $u\mapsto |u(f)|$, $f\in\cS_{p^+}(\R^n)$, respectively, 
$f\in\cS_{p^+\gtrless 0}(\R^n)$. 
\end{definition}
In the following we denote by $\C[p^+,p_\bot]_{p^+}
=S^{-1}\C[p^+,p_\bot]$
the localization (cf.\ \cite{ComAlg}) of the polynomial ring
$\C[p^+,\LCp_\bot]$
by the multiplicative set $S=\{(p^+)^k:k\geq 0\}$. It consists of formal
fractions $\frac{R}{(p^+)^k}$ where $R\in\C[p^+,p_\bot]$, $k\in\N$.
Alternatively, we can describe the elements of $\C[p^+,p_\bot]_{p^+}$
as Laurent polynomials in $p^+$, i.e., each element of $\C[p^+,p_\bot]_{p^+}$
is of the form $T(p^+,p_\bot,1/p^+)$, where
$T(p^+,p_\bot,t)\in\C[p^+,p_\bot,t]$ is a ``usual'' polynomial.
We consider the elements of $\C[p^+,\LCp_\bot]_{p^+}$
as $\C$-valued functions on $\C^n\setminus\{p^+ = 0\}$. Recall that in
algebraic geometry \cite{Hartshorne} $\C[p^+,p_\bot]_{p^+}$
is called the ring of regular functions
on the (distinguished) open set $D(p^+)=\C^n\setminus\{p^+=0\}$.
\begin{remark}
The space $\cS_{p^+}(\R^n)$ can also be defined by the following
seminorms which are equivalent to the seminorms in \eqref{eq:def1}:
\begin{align*}
    ||f||_{Q,\alpha} & =\sup_{\LCbp\in\R^n\setminus\{p^+=0\}}
        |Q(\LCbp)\partial^\alpha f(\LCbp)|,\qquad Q\in \C[p^+,p_\bot]_{p^+},~
        \text{$\alpha$ multi-index},\\
\intertext{or}
    ||f||_{N,\alpha} & =\sup_{\LCbp\in\R^n\setminus\{p^+=0\}}
        \left(\frac{1+|\LCbp|}{|p^+|}\right)^N|\partial^\alpha
        f(\LCbp)|, \qquad N\in\N,~
        \text{$\alpha$ multi-index}.
\end{align*}
\end{remark}
\begin{remark}\label{rem:squeezed_generalized_functions}
1) Let $f\in\cS_{p^+}(\R^n)$. From $||f||_{-k,0,\alpha}=C<\infty$ it follows
$|\partial^\alpha f(\LCbp)|\leq C|(p^+)^k|$ ($k\in\N$, $\alpha$ multi-index).
Hence, as $p^+$ goes to $0$, any partial derivative of $f$
goes faster to $0$ than any power of $p^+$. 
Since $f$ is a $C^\infty$-function, $\partial^\alpha f|_{\{p^+=0\}}=0$.
Moreover, any $f\in\cS_{p^+}(\R^n)$
has the same rapidly decreasing behavior (as $|\LCbp|$ goes to infinity) as
the functions of the Schwartz space $\cS(\R^n)$. More precisely, we have
$\cS_{p^+}(\R^n)\subset\cS(\R^n)$.

2) If
$f$ is a (complex-valued) $C^\infty$-function on $\R^n\setminus\{p^+=0\}$
such that $||f||_{k,\alpha,\beta}<\infty$ for all $k\in\Z$ and all 
multi-indices $\alpha,\beta$ then one can easily verify that $f$ has a 
unique continuous extension to $\{p^+=0\}$ -- necessarily $f(p^+=0)=0$.
This continuous extension of $f$ is
a $C^\infty$-function on $\R^n$, and hence belongs to $\cS_{p^+}(\R^n)$.
In the following we always
consider in such a case the extension (by zero) of $f$ (to $\{p^+=0\}$),
also denoted $f$, without mentioning it explicitly.
\end{remark}

One of the reasons why squeezed generalized functions are so important for us
is the fact that $\cS_{p^+>0}(\R^n)$ (as well as
$\cS_{p^+<0}(\R^n)$) is the $\C$-linear
homeomorphic image of the Schwartz space
$\cS(\R^n)$ under a certain map called the {\em squeezing mapping}.
\begin{definition}\label{def:squeezing_map}
We call the mapping $\nu_{\gtrless 0}:\R_{\gtrless 0}\times\R^{n-1}\to\R^n$
defined by $\nu_{\gtrless 0}=\Omega^{-1}_\pm\circ\kappa^{-1}\circ
\LCOmega_\pm$ the {\em positive/negative squeezing mapping}.
\end{definition}
\begin{proposition}\label{prop:main_isom}
The mapping $f\mapsto j(\nu^*_{\gtrless 0}f)=j(f\circ \nu_{\gtrless 0})$,
where $j(.)$ denotes extension by zero,
defines a linear homeomorphism from
$\cS(\R^n)$ onto $\cS_{p^+\gtrless 0}(\R^n)$.
\end{proposition}
\begin{proof}
cf. \cite{PullJMP}.
\end{proof}
By abuse of language we neglect $j$ in the notation and
write simply $\nu^*_{\gtrless 0}$ instead of $j(\nu^*_{\gtrless 0})$, hence
$\nu^*_{\gtrless 0}:\cS(\R^n)\isomto\cS_{p^+\gtrless 0}(\R^n)$ is an
isomorphism of (complex) topological vector spaces.
\begin{theorem}\label{thm:squeezed_functions}
(i) $\cS_{p^+}(\R^n)\subset\cS(\R^n)$, and the topology of $\cS_{p^+}(\R^n)$
coincides with the subspace topology induced by $\cS(\R^n)$. Moreover,
$\cS_{p^+}(\R^n)$ is a closed subspace of $\cS(\R^n)$.

(ii) $\cS_{p^+}(\R^n)=\cS_{p^+>0}(\R^n)\oplus\cS_{p^+<0}(\R^n)$.

(iii) Consider the following filtration of $\cS(\R^n)$
$$
    \cS(\R^n)\supset p^+\cS(\R^n)\supset (p^+)^2\cS(\R^n)\supset\cdots
$$
Then $\cS_{p^+}(\R^n)=\bigcap_{k\geq 0}(p^+)^k\cS(\R^n)$, or, using categorical
language,
$$
    \cS_{p^+}(\R^n)=\varprojlim_{k\in\N} (p^+)^k\cS(\R^n)
$$
in the category of topological vector spaces.

(iv) If $Q\in\C[p^+,p_\bot]_{p^+}$, $g\in\cS_{p^+}(\R^n)$, and $\alpha$ is
a multi-index, then
$$
    f\mapsto Qf
    ,\qquad f\mapsto gf, \qquad f\mapsto\partial^\alpha f
$$
are continuous, linear mappings from $\cS_{p^+}(\R^n)$ to $\cS_{p^+}(\R^n)$.
\footnote{We consider $Qf$ canonically as a function on $\R^n$
(cf. Remark~\ref{rem:squeezed_generalized_functions})}    
\end{theorem}
\begin{proof}
cf. \cite{PullJMP}.
\end{proof}

In the following we identify $\cS'_{p^+\gtrless 0}(\R^n)$ with the
subspace $\{u\in\cS'_{p^+}(\R^n):u|_{p^+\lessgtr 0}\equiv 0\}$.
Furthermore, we have canonical inclusion mappings 
$\cS_{p^+\gtrless 0}(\R^n)\injto\cS'_{p^+\gtrless 0}(\R^n)$ and
$\cS_{p^+}(\R^n)\injto\cS'_{p^+}(\R^n)$.
\begin{theorem}\label{thm:squeezed_generalized_functions}
(i) $\cS'_{p^+}(\R^n)=\cS'_{p^+>0}(\R^n)\oplus\cS'_{p^+<0}(\R^n)$.

(ii) For each $u\in\cS'_{p^+\gtrless 0}(\R^n)$, there is a sequence
$(u_i)_{i\in\N}$, $u_i\in\cS_{p^+\gtrless 0}(\R^n)$, converging to $u$ in
$\cS'_{\gtrless 0}(\R^n)$. The same holds, if we replace
$\cS_{p^+\gtrless 0}(\R^n)$ (respectively $\cS'_{p^+\gtrless 0}(\R^n)$)
by $\cS_{p^+}(\R^n)$ (respectively $\cS'_{p^+}(\R^n)$).

(iii) The isomorphisms 
$\nu^*_{\gtrless 0}:\cS(\R^n)\to\cS_{p^+\gtrless 0}(\R^n)$ extend (uniquely)
to linear, sequentially continuous, bijective mappings
$\nu^*_{\gtrless 0}:\cS'(\R^n)\to\cS'_{p^+\gtrless 0}(\R^n)$, where the
inverse mappings are also sequentially continuous.
\end{theorem}
\begin{proof}
cf. \cite{PullJMP}.
\end{proof}
%
%
\begin{definition}
(a) We denote by $\cS_{\partial_{x^-}}(\R^n)$ (or briefly 
$\cS_{\partial_-}(\R^n)$)
the complex vector space
$$
    \cS_{\partial_{x^-}}(\R^n):=\bigcap_{m\geq 0}\partial^m_{x^-}\cS(\R^n)=
    \{g:\forall m\geq 0~\exists h\in\cS(\R^n)~g=\partial^m_{x^-}h\}
$$
endowed with the subspace topology induced by $\cS(\R^n)$ which makes
$\cS_{\partial_{x^-}}(\R^n)$ into a locally convex topological vector space.
The dual space $\cS'_{\partial_-}(\R^n)$ is called the space of
{\em tame generalized functions}.

(b) The (continuous) inclusion
mapping $\iota:\cS_{\partial_-}(\R^n)\injto\cS(\R^n)$ induces
a canonical (pullback) mapping $\iota^*:\cS'(\R^n)\to\cS'_{\partial_-}(\R^n)$,
$u\mapsto u^*:=\iota^*u$; we call $u^*$ the {\em relaxation} of $u$.
\end{definition}
Notice that, by the Hahn-Banach Theorem, the mapping $u\mapsto u^*$ is
surjective. The elements of the fibre over $u^*$ are the regularizations
of $u^*$.
\begin{proposition}
The mapping $\pLCFT$ (cf. \eqref{equ:partial_LC-Fourier})
is a linear homeomorphism
from $\cS(\R^n)$ onto $\cS(\R^n)$ which maps the subspace 
$\cS_{\partial_-}(\R^n)$ onto the subspace $\cS_{p^+}(\R^n)$. The canonical
extension of $\pLCFT$ to $\cS'(\R^n)$, also denoted by $\pLCFT$,
maps $\cS'_{\partial_-}(\R^n)$ onto $\cS'_{p^+}(\R^n)$.
\end{proposition}
\begin{corollary}
The space $\cS_{\partial_-}(\R^n)$ is a closed subspace of $\cS(\R^n)$, hence
$\cS_{\partial_-}(\R^n)$ is a Fr\'echet space.
\end{corollary}
\begin{definition}
A function $M\in C^\infty(\R^n\setminus\{p^+=0\},\C)$
is called a {\em multiplicator in $\cS_{p^+}(\R^n)$} if $Mf\in\cS_{p^+}(\R^n)$
for every $f\in\cS_{p^+}(\R^n)$.
\end{definition}
\begin{remark}
As one can easily see, if $M,M_1,M_2$ are multiplicators in 
$\cS_{p^+}(\R^n)$,
$c_1,c_2\in\C$ and $\alpha$ a multi-index, then also $M_1M_2$, 
$c_1M_1+c_2M_2$ and $\partial^\alpha M$ are multiplicators in 
$\cS_{p^+}(\R^n)$. Furthermore, from Theorem~\ref{thm:squeezed_functions}
it follows that any 
$Q\in C[p^+,p_\bot]_{p^+}$ and any $g\in\cS_{p^+}(\R^n)$ is a multiplicator
in $\cS_{p^+}(\R^n)$.
\end{remark}
By standard arguments, as in the case of generalized functions, we obtain:
\begin{proposition}\label{prop:crit_multiplicator}
A function $M\in C^\infty(\R^n\setminus\{p^+=0\},\C)$ is a multiplicator
in $\cS_{p^+}(\R^n)$ if and only if there exists for each multi-index 
$\alpha$ a positive constant $C_\alpha$ and a natural number $N_\alpha$
such that
$$
    |\partial^\alpha M(\LCbp)|\leq C_\alpha\left(
        \frac{1+|\LCbp|}{|p^+|}\right)^{N_\alpha}\qquad
        \text{for all multi-indices $\alpha$ and $\LCbp\in\R^n$}.
$$
\end{proposition}
\begin{corollary}\label{cor:examples_multiplicators}
The functions $\Theta(\pm p^+)$,
$|p^+|^k$ $(k\in\Z)$ are multiplicators in $\cS_{p^+}(\R^n)$,
(hereby $\Theta$ is the Heaviside function defined by $\Theta(p^+)=1$, 
if $p^+>0$, and $\Theta(p^+)=0$, if $p^+<0$.)
\end{corollary}
\begin{definition}
Given a multiplicator $M$ in $\cS_{p^+}(\R^n)$ and a squeezed generalized
function $u\in\cS'_{p^+}(\R^n)$, we define the squeezed generalized
function $Mu\in\cS'_{p^+}(\R^n)$ by $(Mu,f)=(u,Mf)$ for all
$f\in\cS_{p^+}(\R^n)$.
\end{definition}
Our definition of a multiplicator in $\cS_{p^+}(\R^n)$ is closely related
to that of a multiplicator in $\cS(\R^n)$ as defined in \cite{Bog1}. However,
the set of multiplicators in $\cS_{p^+}(\R^n)$ involves much more singular
functions such as $1/p^+$ which is even non-locally integrable.
This feature is crucial in solving the characteristic Cauchy problem.

\subsection{Solutions of the LCKG-equation}
In Section~\ref{sec:rev_Cauchy} we determined the general solution of the
(homogeneous) KG-equation $(\Box+m^2)u=0$
by determining the general solution of the associated division problem
$(p^2-m^2)v=0$. Thereby we introduced the generalized functions
$a_\pm(\bp)\delta(p^2-m^2)\in\cS'(\R^{1+n})$ where
$a_\pm(\bp)\in\cS'(\R^n)$.

In this subsection we would like to determine
the general solution of the (homogeneous)
LCKG-equation $(\LCBox+m^2)\LCu=0$ in the same way. The associated
division problem reads now $(\LCp^2-m^2)\LCv=0$ (recall that
$\LCp^2=2p^+p^- - p_\bot^2$). We start by defining an important class of
generalized functions on $\R^{1+n}$.
\begin{definition}
For each squeezed generalized function $b(\LCbp)\in\cS'_{p^+}(\R^n)$ we
define the generalized functions $b(\LCbp)\delta_\pm(\LCp^2-m^2)$,
$b(\LCbp)\delta(\LCp^2-m^2)\in\cS'(\R^{1+n})$ by
\begin{align*}
    (b(\LCbp)\delta_\pm(\LCp^2-m^2),f(\LCp)) & =
        (b(\LCbp),\frac{\Theta(\pm p^+)}{2|p^+|}f(\LCOmega(\LCbp)))
        \qquad (f\in\cS(\R^{1+n})),\\
    (b(\LCbp)\delta(\LCp^2-m^2),f(\LCp)) & =
        (b(\LCbp),\frac{1}{2|p^+|}f(\LCOmega(\LCbp)))
        \qquad (f\in\cS(\R^{1+n})),
\end{align*}
where $\LCOmega(\LCbp)=(\LCbp,\LComega(\LCbp))$ and
$\LComega(\LCbp)=\frac{p_\bot^2+m^2}{2p^+}$, 
$p_\bot^2=\sum_{i=1}^{n-1}(p^i)^2$.
\end{definition}
\begin{remark}
The above definitions make sense since
$\Theta(\pm p^+)$ and $|p^+|^{-1}$ are multiplicators in $\cS_{p^+}(\R^n)$,
by Corollary~\ref{cor:examples_multiplicators}, and since
$f\mapsto f(\LCbp,\LComega(\LCbp))$ is a 
$\C$-linear, continuous mapping from $\cS(\R^{1+n})$
to $\cS_{p^+}(\R^n)$. This follows easily from
$\LCOmega_\pm=\kappa\circ\Omega_\pm\circ\nu_{\gtrless 0}$ and the fact
that $f\mapsto f\circ\kappa\circ\Omega_\pm$ is a linear, continuous mapping
from $\cS(\R^{1+n})$ to $\cS(\R^n)$, and $g\mapsto g\circ\nu_{\gtrless 0}$
is a linear, continuous mapping from $\cS(\R^n)$ to $\cS_{p^+}(\R^n)$
(cf. Proposition~\ref{prop:main_isom}).
\end{remark}
\begin{remark}
a) If $b(\LCbp)\in\cS'_{p^+}(\R^n)$, $b_+(\LCbp)\in\cS'_{p^+>0}(\R^n)$ and
$b_-(\LCbp)\in\cS'_{p^+<0}(\R^n)$ are given such that $b=b_+ + b_-$ 
holds in $\cS'_{p^+}(\R^n)$ then
$$
    b(\LCbp)\delta(\LCp^2-m^2) = b_+(\LCbp)\delta_+(\LCp^2-m^2) +
        b_-(\LCbp)\delta_-(\LCp^2-m^2).
$$

b) Let $\LCGamma_m=\{\LCp\in\R^{1+n}:\LCp^2-m^2=0\}$ be the LC-mass
hyperboloid, and $\LCGamma^\pm_m=\LCGamma_m\cap\{\pm p^->0\}$. Then, if
$\kappa:\M^{1+n}\to\LC^{1+n}$ is the transformation to LC-coordinates,
$\kappa(\Gamma^\pm_m)=\LCGamma^\pm_m$. In contrast to the Minkowski-case,
cf. Remark~\ref{rem:r1}\,a), the projection map $\LCGamma_m\to\R^n$, 
$\LCp=(\LCbp,p^-)\mapsto\LCbp$ induces a homeomorphism from $\LCGamma_m$
onto $\R^n\setminus\{p^+=0\}$ whose inverse mapping is 
$\R^n\setminus\{p^+=0\}\to\LCGamma_m$, $\LCbp\mapsto\LCOmega(\LCbp)$. 
We denote by $\LCOmega_\pm$ the restriction of $\LCOmega$ to $\{\pm p^+>0\}$, then
$\LCOmega_\pm$ maps $\{\pm p^+>0\}$ onto $\LCGamma^\pm_m$.

c) If $b(\LCbp)\subset\cS_{p^+}(\R^n)(\subset\cS'_{p^+}(\R^n))$ then
$$
    (b(\LCbp)\delta_\pm(\LCp^2-m^2),f(\LCbp))=
     \int_{\LCGamma^\pm_m}\frac{(b\circ\LCOmega^{-1})(\LCp) f(\LCbp)}
     {|\nabla\LCQ(\LCp)|}d\LCS(\LCp),
$$
where $\LCQ(\LCp)=\LCp^2-m^2$ and $d\LCS$ is the canonical surface measure
on $\LCGamma_m$.

d) The support of
$b(\LCbp)\delta_\pm(\LCp^2-m^2)$ is contained in $\LCGamma^\pm_m$.
\end{remark}
\begin{lemma}\label{lem:sequ_cont}
The mappings $\cS'_{p^+}(\R^n)\to\cS'(\R^{1+n})$,
$b(\LCbp)\mapsto b(\LCbp)\delta_\pm(\LCp^2-m^2)$ and
$b(\LCbp)\mapsto b(\LCbp)\delta(\LCp^2-m^2)$
are $\C$-linear and sequentially continuous.
\end{lemma}
\begin{proof}
This follows immediately from the definitions of the generalized functions
$b(\LCbp)\delta_\pm(\LCp^2-m^2)$ and $b(\LCbp)\delta(\LCp^2-m^2)$.
\end{proof}
The next proposition is essential for the sequel since it gives us the proper
transformation law between the generalized functions
$a(\bp)\delta_\pm(p^2-m^2)$ $(a(\bp)\in\cS'(\R^n))$ and 
$b(\LCbp)\delta_\pm(\LCbp^2-m^2)$ $(b(\LCbp)\in\cS'_{p^+}(\R^n))$.
Notice that $\nu^*_{\gtrless 0}$ is an isomorphism from
$\cS'(\R^n)$ onto $\cS'_{p^+\gtrless 0}(\R^n)$.
\begin{proposition}\label{prop:transformation2}
Let $a(\bp)\in\cS'(\R^n)$ be a generalized function and let 
$\kappa:\R^{1+n}\to\R^{1+n}$ be the transformation to LC-coordinates. Then
$$
    a(\bp)\delta_\pm(p^2-m^2)\circ\kappa^{-1}=
        (\nu^*_{\gtrless 0}a)(\LCbp)\delta_\pm(\LCp^2-m^2).
$$
\end{proposition}
\begin{proof}
Both sides of the above equation depend (sequentially) continuously on 
$a(\bp)\in\cS'(\R^n)$.
Hence it is enough to show equality, if $a(\bp)$ is from the dense subspace
$\cS(\R^n)$ of $\cS'(\R^n)$. So, assume $a(\bp)\in\cS(\R^n)$ and let 
$f\in\cS(\R^{1+n})$. Then
\begin{align*}
    (a(\bp) & \delta_\pm(p^2-m^2)\circ\kappa^{-1},f) =
    \int\frac{d^n\bp}{2\omega(\bp)}a(\bp)(f\circ\kappa\circ\Omega_\pm)(\bp)=\\
    & =\int_{\Gamma^\pm_m}
    \frac{(a\circ\Omega^{-1}_\pm)(f\circ\kappa)}{|\nabla Q|}dS=
    \int_{\LCGamma^\pm_m}
    \frac{(a\circ\Omega^{-1}_\pm\circ\kappa^{-1})f}{|\nabla\LCQ|}d\LCS=\\
    & = \int_{p^+\gtrless 0}\frac{d^n\LCbp}{2|p^+|}
    (a\circ\Omega^{-1}_\pm\circ\kappa^{-1}\circ\LCOmega_\pm)(\LCbp)
    (f\circ\LCOmega_\pm)(\LCbp)=\\
    & = \left((\nu^*_{\gtrless 0}a)(\LCbp)\delta_\pm(\LCp^2-m^2),f\right).
\end{align*}
\end{proof}
\begin{corollary}\label{cor:injectivity}
Let $b_\pm(\LCbp)\in\cS'_{p^+\gtrless 0}(\R^n)$ and
$b(\LCbp)\in\cS'_{p^+}(\R^n)$.

(i) $b_\pm(\LCbp)\delta_\pm(\LCp^2-m^2)=0$
if and only if $b_\pm(\LCbp)=0$.

(ii) $b(\LCbp)\delta(\LCp^2-m^2)=0$ if and only if $b(\LCbp)=0$.
\end{corollary}
\begin{proof}
(i) This can be proven directly as in Lemma~\ref{lem:inj}(i), however,
we will reduce it to Lemma~\ref{lem:inj}. Assume 
$\LCu=b_\pm(\LCbp)\delta_\pm(\LCp^2-m^2)=0$, then also 
$u=\LCu\circ\kappa=0$. By Proposition~\ref{prop:transformation2},
$u=a_\pm(\bp)\delta_\pm(p^2-m^2)$ with 
$a_\pm=(\nu^*_{\gtrless 0})^{-1}b_\pm\in\cS'(\R^n)$. Now, 
by Lemma~\ref{lem:inj}\,(i), $a_\pm(\bp)=0$ and hence $b_\pm(\LCbp)=0$.

(ii) Assume $b(\LCbp)\delta(\LCp^2-m^2)=0$. By 
Theorem~\ref{thm:squeezed_generalized_functions}\,(i) there are unique
$b_\pm(\LCbp)\in\cS'_{p^+\gtrless 0}(\R^n)$ such that
$b(\LCbp)=b_+(\LCbp) + b_-(\LCbp)$. Since from
$0=b(\LCbp)\delta(p^2-m^2)=b_+(\LCbp)\delta_+(\LCp^2-m^2) + 
b_-(\LCbp)\delta_-(\LCp^2-m^2)$ we obtain $b_+(\LCbp)\delta_+(\LCp^2-m^2)=
-b_-(\LCbp)\delta_-(\LCp^2-m^2)$ and since these distributions have disjoint
supports, the assertion follows from (i).
\end{proof}
\begin{remark}
Corollary~\ref{cor:injectivity} shows that 
$\cS'_{p^+\gtrless 0}(\R^n)\to\cS'(\R^{1+n})$, 
$b(\LCbp)\mapsto b(\LCbp)\delta_\pm(\LCp^2-m^2)$ and 
$\cS'_{p^+}(\R^n)\to\cS'(\R^{1+n})$, 
$b(\LCbp)\mapsto b(\LCbp)\delta(\LCp^2-m^2)$ are injective maps. 
\end{remark}
\begin{corollary}\label{cor:LC_general_solution}
The general solution of the division problem $(\LCp^2-m^2)\LCu=0$ in
$\cS'(\R^{1+n})$ is given by
\begin{equation}\label{equ:gen_sol}
    \LCu=b_+(\LCbp)\delta_+(\LCp^2-m^2)+b_-(\LCbp)\delta_-(\LCp^2-m^2)
\end{equation}
with $b_+(\LCbp)\in\cS'_{p^+ > 0}(\R^n)$ and
$b_-(\LCbp)\in\cS'_{p^+ < 0}(\R^n)$, or, equivalently, by
$$
    \LCu = b(\LCbp)\delta(\LCp^2-m^2)
$$
with $b\in\cS_{p^+}'(\R^n)$.
\end{corollary}
\begin{proof}
The mapping $u\mapsto \LCu=u\circ\kappa^{-1}$ defines a 1-1 correspondence
between the solutions of $(p^2-m^2)u=0$ and the solutions of 
$(\LCp^2-m^2)\LCu=0$. The general solution of $(p^2-m^2)u=0$ is of the
form $u=a_+(\bp)\delta_+(p^2-m^2)+a_-(\bp)\delta_-(p^2-m^2)$ with
$a_\pm(\bp)\in\cS'(\R^n)$, hence the assertion follows from
Proposition~\ref{prop:transformation2}. 
\end{proof}
\begin{remark}
The general solution of the division problem $(\LCp^2-m^2)\LCu=1$ in
$\cS'(\R^{1+n})$ is given by
$$
    \LCu=\PV\frac{1}{\LCp^2-m^2}+
    b_+(\LCbp)\delta_+(\LCp^2-m^2)+b_-(\LCbp)\delta_-(\LCp^2-m^2)
$$
where $b_\pm(\LCbp)\in\cS'_{p^+\gtrless 0}(\R^n)$, or, equivalently, by
$$
    \LCu=\PV\frac{1}{\LCp^2-m^2}+b(\LCbp)\delta(\LCp^2-m^2)
$$
where $b(\LCbp)\in\cS'_{p^+}(\R^n)$ where $\PV$ denotes the principal value.
\end{remark}
\begin{corollary}\label{cor:sol_representation}
Let $u\in\cS'(\R^{1+n})$ be a solution of the LCKG-equation $(\LCBox+m^2)u=0$.
Then there
is a unique $b(\LCbp)\in\cS'_{p^+}(\R^n)$ such that 
$u=\cF_\LC^{-1}(b(\LCbp)\delta(\LCp^2-m^2))$.
\end{corollary}
\begin{proof}
Since $u\in\cS'(\R^{1+n})$ is a solution of the LCKG-equation, 
$\cF_\LC(u)=u^\hatL$ is a solution of the division problem
$(\LCp^2-m^2)u^\hatL=0$. The existence follows now from 
Corollary~\ref{cor:LC_general_solution} and the uniqueness from
Corollary~\ref{cor:injectivity}.
\end{proof}
\subsection{The tame restriction of a generalized function}
There are several equivalent ways
to define canonically the restriction of a distribution $u$ on $\R^{1+n}$ to
a hyperplane $\Sigma\subset\R^{1+n}$. One approach  uses an extended
construction of the pullback of a distribution to define the
restriction $u|_\Sigma$ as the pullback $\iota^*u$, where 
$\iota:\Sigma\injto\R^{1+n}$ is the inclusion mapping.
By this, the restriction is canonically definable
if and only if the normal bundle of (the submanifold) $\Sigma$
doesn't intersect the wave front set of $u$ (cf. \cite{Hoer1}, 8.2).
By another way one uses the definition of dependence of a distribution on 
a parameter and defines the restriction by just fixing the parameter (cf.,
e.g., \cite{Bog1} 2.6.). We will define the tame restriction
of a generalized function by using the second method, where connections with
pullbacks and wave front sets will be elaborated in a further paper
\cite{Pull_future1}.
\begin{definition}
A tame generalized function $u=u(x^+,\LCbx)\in\cS'_{\partial_-}(\R^{1+n})$
is called {\em $C^\infty$-dependent} on $x^+\in\Omega$ as a
parameter ($\Omega\subset\R$ open) if
there exists a family $(u_{x^+})_{x^+\in\Omega}$ in $\cS'_{\partial_-}(\R^n)$,
$u_{x^+}=u_{x^+}(\LCbx)$, such that
$$
   (u(x^+,\LCbx),f(x^+)\otimes g(\LCbx))=\int_\Omega dx^+ (u_{x^+},g)
   f(x^+)
$$
for all $f\in\cD(\Omega)$, $g\in\cS_{\partial_-}(\R^n)$, and such that
$$
	\Omega\to\C,~x^+\mapsto (u_{x^+},g)
$$
is $C^\infty$ for all $g\in\LCcS(\R^n)$.
\end{definition}
\begin{remark}
Obviously, the family $(u_{x^+})_{x^+\in\Omega}$ is uniquely determined
by $u$.
\end{remark}
Since the LCKG-operator $\LCBox+m^2$ is not hypoelliptic with respect to
$x^+$ there are solutions $u\in\cD'(\R^{1+n})$ of the
LCKG-equation which are not $C^\infty$-dependent on $x^+$ as a parameter.
For instance, the positive/negative frequency Pauli-Jordan functions
$D^{(\pm)}=\frac{\pm 1}{i(2\pi)^3}\cF_\M(\delta_\pm(p^2-m^2))\in\cS'(\R^4)$ do not 
have a canonical restriction to $\{x^0+x^3=0\}$. This can be seen by considering the
wave front set of $D^{(\pm)}$ (see \cite{UllWer04a}).
Now the following proposition shows that we can retrieve
parameter dependence if we consider the relaxation $u^*$ instead of $u$.
\begin{proposition}\label{prop:parameter_dependance}
Assume $b(\LCbp)\in\cS'_{p^+}(\R^n)$ and let
$u(\LCx)=\cF_\LC^{-1}(b(\LCbp)\delta(\LCp^2-m^2))\in\cS'(\R^{1+n})$.
Then $u^*(\LCx)=u^*(x^+,\LCbx)$
is $C^\infty$-dependent on $x^+$ as a parameter,
and for all $x^+\in\R$
$$
    u^*_{x^+} = \frac{1}{4\pi}(\pLCFT)^{-1}\left(
    \frac{e^{-i\LComega(\LCbp)x^+}}{|p^+|}b(\LCbp)\right).
$$
\end{proposition}
\begin{proof}
For each $x^+\in\R$ we set
$u^*_{x^+}=\frac{1}{4\pi}(\pLCFT)^{-1}
\left(\frac{e^{-i\LComega(\LCbp)x^+}}{2|p^+|}b(\LCbp)\right).$
Since, by Proposition~\ref{prop:crit_multiplicator},
$|p^+|^{-1}e^{-i\LComega(\LCbp)x^+}$ is a multiplicator in
$\cS_{p^+}(\R^n)$,
$u_{x^+}\in\cS'_{\partial_-}(\R^n)$ is well-defined for each $x^+\in\R$.
Moreover, for every $g\in\cS_{p^+}(\R^n)$, 
$$
	\R\to\C,~
	x^+\mapsto (b(\LCbp),2^{-1}|p^+|^{-1}e^{-i\LComega(\LCbp)x^+}g(\LCbp))
$$
is a $C^\infty$-mapping and thus
$\R\to\C$, $x^+\mapsto (u_{x^+},g)$ is a $C^\infty$-mapping, too.
It remains to show, that
\begin{equation}\label{equ1:prop:parameter_dependence}
    (u(x^+,\LCbx),f(x^+)g(\LCbx))=\int(u_{x^+},g)f(x^+)dx^+
\end{equation}
for each $f(x^+)\in\cS(\R)$ and $g(\LCbx)\in\cS_{\partial_-}(\R^n)$.

\textbf{Case 1:}
Firstly, we assume that $b(\LCbp)\in\cS_{p^+}(\R^n)(\subset\cS'_{p^+}(\R^n))$.
By an easy computation we obtain
\begin{align*}
    (u^*,f\otimes g) & = (b(\LCbp)\delta(\LCp^2-m^2),\cF^{-1}_\LC(f\otimes g))\\
        & = \frac{1}{(2\pi)^{n+1}}
            \int dx^+ f(x^+)\int d^n\LCbx\,g(\LCbx)\int
            \frac{d^n\LCbp}{2|p^+|}b(\LCbp)
            e^{-i(\LComega(\LCbp)x^+ + p^+x^- - p_\bot\cdot x_\bot)}\\
        & = \frac{1}{2\pi}\int dx^+ f(x^+)
            \int\frac{d^n\LCbp}{2|p^+|}b(\LCbp) e^{-i\LComega(\LCbp)x^+}
            ((\pLCFT)^{-1}g)(\LCbp)\\
        & = \int dx^+ (u^*_{x^+},g)f(x^+)
\end{align*}
for $f\in\cS(\R)$, $g\in\cS_{\partial_-}(\R^n)$.

\textbf{Case 2:} Now we consider the general case 
$b(\LCbp)\in\cS'_{p^+}(\R^n)$. By 
Theorem~\ref{thm:squeezed_generalized_functions} 
there is a sequence
$(b^{(m)}(\LCbp))$ in $\cS_{p^+}(\R^n)$ converging to 
$b(\LCbp)$ in $\cS'_{p^+}(\R^n)$. Define $u^{(m)}\in\cS'(\R^{1+n})$ by
$(u^{(m)})^\hatL=b^{(m)}(\LCbp)\delta(\LCp^2-m^2)$. By 
Lemma~\ref{lem:sequ_cont} the sequence $(u^{(m)})$ converges to
$u$ in $\cS'(\R^{1+n})$ and, by construction, $(u^{(m)*}_{x^+})$
converges to $u^*_{x^+}$ in $\cS'_{\partial_-}(\R^n)$. By the first case
we have
\begin{equation}\label{equ2:prop:parameter_dependence}
    (u^*(x^+,\LCbx),f(x^+)g(\LCbx))=
        \lim_{m\to\infty}\int(u^{(m)*}_{x^+},g)f(x^+)dx^+
\end{equation}
for $f\in\cS(\R)$, $g\in\cS_{\partial_-}(\R^n)$. Since
$|(u^{(m)*}_{x^+},g)|\leq\int\frac{d^n\LCbp}{2|p^+|}|b^{(m)}(\LCbp)
h(\LCbp)|=C<\infty$ ($h=(\pLCFT)^{-1}g$),
where the constant $C$ does not depend on $x^+$, the
right-hand side of \eqref{equ2:prop:parameter_dependence} equals
$\int (u_{x^+},g)f(x^+)dx^+$, hence we have shown 
\eqref{equ1:prop:parameter_dependence}.
\end{proof}
\begin{remark}
To see in the above proof that, for each $g\in\cS_{p^+}(\R^n)$, the
mapping $\R\to\C$, $x^+\mapsto 
(b(\LCbp),2^{-1}|p^+|^{-1}e^{-i\LComega(\LCbp)x^+}g(\LCbp))$ 
is $C^\infty$, notice
that for any multiplicator $M$ in $\cS_{p^+}(\R^n)$ the sequences
$$
    e^{-i\LComega(\LCbp)(x^+ + h)}M(\LCbp)g(\LCbp)\to
	e^{-i\LComega(\LCbp)x^+}M(\LCbp)g(\LCbp)\qquad (h\to 0)
$$
and
$$
 \frac{1}{h}\left(e^{-i\LComega(\LCbp)(x^+ + h)}
	-e^{-i\LComega(\LCbp)x^+}\right)M(\LCbp)g(\LCbp)
	\to
	-i\LComega(\LCbp)e^{-i\LComega(\LCbp)x^+}M(\LCbp)g(\LCbp)\qquad (h\to 0)
$$ 
converge in $\cS_{p^+}(\R^n)$ for every $x^+\in\R$.
\end{remark}
\begin{corollary}
The relaxation $u^*(x^+,\LCbx)$
of any solution $u(x^+,\LCbx)\in\cS'(\R^{1+n})$ of the
LCKG-equation $(\LCBox+m^2)u=0$ is $C^\infty$-dependent on $x^+\in\R$
as a parameter.
\end{corollary}
\begin{proof}
Let $u=u(x^+,\LCbx)\in\cS'(\R^{1+n})$ be a solution of the LCKG-equation.
By Corollary~\ref{cor:sol_representation} there is a (unique) 
$b(\LCbp)\in\cS'_{p^+}(\R^n)$ such that $u^\hatL=b(\LCbp)\delta(\LCp^2-m^2)$.
Now apply Proposition~\ref{prop:parameter_dependance}.
\end{proof}
\begin{definition}
A generalized function $u(x^+,\LCbp)\in\cS'(\R^{1+n})$
admits a {\em tame restriction to} $\{x^+=0\}$ if there is an
open neighbourhood $\Omega$ of $0\in\R$ such that the relaxation 
$u^*(x^+,\LCbx)$ of $u$ is $C^\infty$-dependent
on $x^+\in\Omega$ as a parameter.
If $(u^*_{x^+})_{x^+\in\Omega}$ is the corresponding family in 
$\cS'_{\partial_-}(\R^n)$, we set $u\tres_{x^+=0}=u^*_0$ and call
$u\tres_{x^+=0}$ the {\em tame restriction of $u$ to $\{x^+=0\}$}.
\end{definition}
\begin{corollary}
Any solution $u\in\cS'(\R^{1+n})$ of the LCKG-equation admits a tame
restriction to $\{x^+=0\}$.
\end{corollary}
%
%
\section{The characteristic Cauchy problem}
%
In the preceding section we have treated the problem of restricting solutions
$u(\LCx)\in\cS'(\R^{1+n})$ of the LCKG-equation $(\LCBox+m^2)u=0$
to $\{x^+=0\}$. As already mentioned, there exist solutions
$u\in\cS'(\R^n)$ which do not have a canonical restriction to $\{x^+=0\}$
in $\cS'(\R^n)$. We solved this
problem by introducing the space $\cS_{\partial_-}(\R^n)$ along with
its dual space $\cS'_{\partial_-}(\R^n)$. Since $\cS_{\partial_-}(\R^n)$
is a subspace of $\cS(\R^n)$, the elements of $\cS'_{\partial_-}(\R^n)$
may have more singularities than those in $\cS'(\R^n)$.
Now any solution of the LCKG-equation in $\cS'_{\partial_-}(\R^{1+n})$
is $C^\infty$-dependent on $x^+$ as a parameter, and yet has a restriction
to $\{x^+=0\}$ in $\cS'_{\partial_-}(\R^n)$.
 Hence it is possible to assign to each solution $u\in\cS'(\R^{1+n})$ of the LCKG-equation
a tame restriction $u\tres_{x^+=0}$
to $\{x^+=0\}$ in $\cS'_{\partial_-}(\R^n)$ by considering $u$ canonically
as an element $u^*$ of $\cS'_{\partial_-}(\R^{1+n})$ -- we call $u^*$
the relaxation of $u$.
This makes possible to consider the following characteristic Cauchy problem
of the LCKG-equation:
\begin{equation}\label{eq:char_Cauchy}
    (\LCBox +m^2)u=0,\quad\qquad u\tres_{x^+=0}=u_0,
        \qquad\qquad
        (\LCBox=2\partial_+\partial_--\sum_{i=1}^{n-1}\partial^2_i)
\end{equation}
where $u=\cS'(\R^{1+n})$ and $u_0\in\cS'_{\partial_-}(\R^n)$.
\subsection{Existence and uniqueness}\label{sub:well-posedness}
Since $\cS_{\partial_-}(\R^{1+n})$ is not dense in
$\cS(\R^{1+n})$, the induced (relaxation) map
$\cS'(\R^{1+n})\to\cS'_{\partial_-}(\R^{1+n})$, $u\mapsto u^*$
is by no means injective. However, we will show in this subsection that
this is true on the subspace of all $u\in\cS'(\R^{1+n})$ which are
solutions of the LCKG-equation. Moreover, such $u$ are uniquely determined by
their tame restriction $u\tres_{x^+=0}\in\cS'_{\partial_-}(\R^n)$.
Furthermore, we will show that for any $u_0\in\cS'_{\partial_-}(\R^n)$ there exists 
a solution $u\in\cS'(\R^{1+n})$ of the LCKG-equation such that 
$u\tres_{x^+=0}=u_0$.
\begin{lemma}\label{lem:main}
Assume $u\in\cS'(\R^{1+n})$ is a solution of the LCKG-equation
$(\LCBox+m^2)u=0$ and let $u_0=u\tres_{x^+=0}\in\cS'_{\partial_-}(\R^n)$ be
the tame restriction of $u$ to $\{x^+=0\}$. Then
\begin{equation}\label{lem:main_eq1}
    u^\hatL=4\pi|p^+|\pLCFT(u_0)\,\delta(\LCp^2-m^2).
\end{equation}
\end{lemma}
\begin{proof}
By Corollary~\ref{cor:sol_representation} there is a unique
$b(\LCbp)\in\cS'_{p^+}(\R^n)$ such that $u^\hatL=b(\LCbp)\delta(\LCp^2-m^2)$,
and by Proposition~\ref{prop:parameter_dependance} we
obtain $\pLCFT(u_0)=\frac{1}{4\pi|p^+|}b(\LCbp)$.
\end{proof}
\begin{theorem}\label{thm:main}
(i) Suppose $u\in\cS'(\R^{1+n})$ is a solution of the LCKG-equation
and $u\tres_{x^+=0}=0$ in $\cS'_{\partial_-}(\R^n)$, then $u=0$.

(ii) For each $u_0\in\cS'_{\partial_-}(\R^n)$ there exists a (unique)
solution $u\in\cS'(\R^{1+n})$ of the LCKG-equation such that $u\tres_{x^+=0}=u_0$.
\end{theorem}
\begin{proof}
(i) Since $u$ is a solution of the LCKG-equation,
$u^\hatL=4\pi|p^+|\pLCFT(u_0)\,\delta(\LCp^2-m^2)$,
by Lemma~\ref{lem:main}, where
$u_0=u\tres_{x^+=0}$. By assumption $u_0=0$, hence $u=0$.

(ii) Assume $u_0\in\cS'_{\partial_-}(\R^{n-1})$. Define $u\in\cS'(\R^n)$
by $u^\hatL=4\pi|p^+|\pLCFT(u_0)\,\delta(\LCp^2-m^2)$.
Then $(\LCp^2-m^2)u^\hatL=0$
and thus $(\LCBox+m^2)u=0$. By Lemma~\ref{lem:main} and
Corollary~\ref{cor:sol_representation} we obtain
$4\pi|p^+|\pLCFT(u_0)=4\pi|p^+|\pLCFT(u\tres_{x^+=0})$, and thus
$u\tres_{x^+=0}=u_0$ -- notice that $1/|p^+|$ is a multiplicator in $\cS'_{p^+}(\R^n)$.
\end{proof}
The uniqueness in Theorem~\ref{thm:main} implies the following surprising result.
\begin{corollary}
Let $v\in\cS'(\R^{1+n})$ be an arbitrary generalized function. Then any
solution $u\in\cS'(\R^{1+n})$ of the inhomogeneous LCKG-equation 
\begin{equation}\label{eq:inhom_relax}
	(\LCBox+m^2)u=v
\end{equation}
 is uniquely determined by its relaxation $u^*\in\LCcS'(\R^{1+n})$.
\end{corollary}
\begin{proof}
If $v=0$ then the statement follows immediately from the uniqueness in 
Theorem~\ref{thm:main}. Now, in the general case, let $u_1,u_2\in\cS'(\R^{1+n})$ be
two solutions of \eqref{eq:inhom_relax} such that $u^*_1=u^*_2$. Since $u_1 - u_2$ is
a solution of the homogeneous LCKG-equation and $(u_1 - u_2)^*=u^*_1 - u^*_2=0$,
we obtain $u_1 - u_2 =0$ from the first case.
\end{proof}
\subsection{Connection between the characteristic and the non-characteristic
Cauchy problem}\label{sub:relation}
Let $u\in\cS'(\R^{1+n})$ be a solution of the Klein-Gordon equation;
then surely $\LCu=u\circ\kappa^{-1}$ solves the LC-Klein-Gordon equation. Hence we can
consider the several initial data $u_0=u|_{x^0=0}$,
$u_1=\partial_0u|_{x^0=0}\in\cS'(\R^n)$ and
$\LCu_0=\LCu\tres_{x^+=0}\in\cS'_{\partial_-}(\R^n)$. Since we have explicit
formulas relating the initial data $u_0,u_1$ to $u$
(cf.\ Propositions~\ref{prop:general_solution1},~
\ref{prop:main_non-char_Cauchy}) and the initial data
$\LCu_0$ to $\LCu$ (cf.\ \eqref{lem:main_eq1}) we can make use of the
transformation law of Proposition~\ref{prop:transformation2} to obtain
a transformation law between these initial data. Especially,
we obtain again a proof of existence and uniqueness of the characteristic Cauchy problem 
\eqref{eq:char_Cauchy}.
\begin{theorem}
Suppose $u\in\cS'(\R^{1+n})$ is a solution of the Klein-Gordon equation and
let $\LCu=u\circ\kappa^{-1}$. Then, between $u_0=u|_{x^0=0}$,
$u_1=\partial_0u|_{x^0=0}$ and $\LCu_0=\LCu\tres_{x^+=0}$, the
following transformation laws hold:
\begin{align}
    \LChat{\LCu_0} & = \frac{1}{2|p^+|}\left(
        \nu^*_{>0}(\omega u_0^\wedge + i u_1^\wedge) +
        \nu^*_{<0}(\omega u_0^\wedge - i u_1^\wedge)
        \right), \label{equ:e1}
\intertext{and}
         u_0^\wedge & = \frac{1}{\omega}\left(
            \mu^*_{>0}(|p^+|\Theta(p^+)\LChat{\LCu_0}) +
            \mu^*_{<0}(|p^+|\Theta(-p^+)\LChat{\LCu_0})\right),
            \label{equ:e2}\\
         u_1^\wedge & = \frac{1}{i}\left(
            \mu^*_{>0}(|p^+|\Theta(p^+)\LChat{\LCu_0}) -
            \mu^*_{<0}(|p^+|\Theta(-p^+)\LChat{\LCu_0})\right),
            \label{equ:e3}
\end{align}
where $\mu_{\gtrless 0}=\nu_{\gtrless 0}^{-1}:
\R^n\to\R^n\setminus\{\mp p^+\geq 0\}$, $\bp\mapsto\LCbp=(p^+,p_\bot)=
(\frac{1}{\sqrt 2}(p^n\pm\omega(\bp)),p^1,\ldots,p^{n-1})$,
and $\omega=\omega(\bp)=\sqrt{\bp^2 + m^2}$.
\end{theorem}
\begin{proof}
Let $u\in\cS'(\R^{1+n})$ be a solution of the Klein-Gordon equation, and
let $\LCu=u\circ\kappa^{-1}$. Then, by Propositions~
\ref{prop:general_solution1},~\ref{prop:main_non-char_Cauchy},
$u^\hatM=a_0(\bp)\delta_+(p^2-m^2)+a_1(\bp)\delta_-(p^2-m^2)$, where
$a_0=2\pi(\omega u_0^\wedge + i u_1^\wedge)\in\cS'(\R^n)$ and
$a_1=2\pi(\omega u_0^\wedge - i u_1^\wedge)\in\cS'(\R^n)$. On
the other hand, by Lemma~\ref{lem:main}, $u^\hatL=b(\LCbp)\delta(\LCp^2-m^2)$,
where $b(\LCbp)=4\pi|p^+|\LChat{\LCu_0}(\LCbp)$. We write
$b(\LCbp)\delta(\LCp^2-m^2)=b_0(\LCbp)\delta_+(\LCp^2-m^2)+
b_1(\LCbp)\delta_-(\LCp^2-m^2)$ with $b_0(\LCbp)=\Theta(p^+)b(\LCbp)$ and
$b_1(\LCbp)=\Theta(-p^+)b(\LCbp)$. Since $\LCu^\hatL=u^\hatM\circ\kappa^{-1}$ we
obtain $b_0=\nu^*_{>0}(a_0)$ and $b_1=\nu^*_{<0}(a_1)$ by
Proposition~\ref{prop:transformation2} and Corollary~\ref{cor:injectivity},
and hence
\begin{align*}
    2|p^+|\Theta(p^+)\LChat{\LCu_0} & =
        \nu^*_{>0}(\omega u_0^\wedge + iu_1^\wedge),\\
    2|p^+|\Theta(-p^+)\LChat{\LCu_0} & =
        \nu^*_{<0}(\omega u_0^\wedge - iu_1^\wedge),
\end{align*}
from which the transformation laws easily follow.
\end{proof}
\begin{corollary}
Suppose the initial data $u_0,u_1\in\cS'(\R^n)$
and $\LCu_0\in\cS'_{p^+}(\R^n)$ fulfill
\eqref{equ:e1} or, equivalently,
\eqref{equ:e2} and \eqref{equ:e3}. Then $u\in\cS'(\R^{1+n})$
is a solution of the (non-characteristic) Cauchy problem
$$
    (\Box+m^2)u=0,\qquad u|_{x^0=0}=u_0,\qquad \partial_0 u|_{x^0=0}=u_1
$$
if and only if $\LCu=u\circ\kappa^{-1}\in\cS'(\R^{1+n})$ is a solution of the
(characteristic) Cauchy problem
$$
    (\LCBox + m^2)\LCu=0,\qquad \LCu\tres_{x^+=0}=\LCu_0.
$$
In this case, $u$ (respectively $\LCu$) is uniquely determined by
the initial data $u_0$, $u_1$ (respectively $\LCu_0$).
\end{corollary}
\begin{example}
Consider the Pauli-Jordan function
$$
	D_m(x)=\frac{1}{i(2\pi)^3}\int d^4p\,
	\epsilon(p^0)\delta(p^2-m^2) e^{i\Mbil{p}{x}}
	\in\cS'(\R^4).
$$
which is a fundamental solution of the Cauchy problem of the Klein-Gordon equation, i.e.,
$$
	(\Box+m^2)D_m=0 \quad\text{and}\quad
	D_m|_{x^0=0}=0,\quad\partial_0D_m|_{x^0=0}=\delta(\bx)
$$
Transforming $D_m$ to light cone coordinates we obtain $\LCD_m=D_m\circ\kappa^{-1}$ 
which is a solution of the LCKG-equation, i.e., $(\LCBox+m^2)\LCD_m=0$.
Now we can use the transformation law \eqref{equ:e1} to compute the tame
restriction $\LCD_m\tres_{x^+=0}$ directly from the restrictions $D_m|_{x^0=0}$ and 
$\partial_0 D_m|_{x^0=0}$. Since the Fourier transform
of $\delta(\bx)$ is the constant function $1\in\cS'(\R^3)$ we get
$$
	\LChat{\left(\LCD_m\tres_{x^+=0}\right)}=\frac{i}{2|p^+|}\left(
		\nu^*_{>0}1 - \nu^*_{<0} 1 \right).
$$
and hence for all $g\in\cS_{p^+}(\R^3)$
$$
	\left(\LChat{\left(\LCD_m\tres_{x^+=0}\right)},g\right)=
		i\int_{p^+>0}\frac{d^3\LCbp}{2p^+}
		\left( g(\LCbp) - g(-\LCbp) \right).
$$
Applying the Fourier inversion formula, we obtain
\begin{equation}\label{equ:eq4}
	\left(\LCD_m\tres_{x^+=0},f\right)=\frac{i}{(2\pi)^3}\int_{p^+>0}
		\frac{d^3\LCbp}{2p^+}
		\left(\LChat f(-\LCbp) - \LChat f(\LCbp)\right).
\end{equation}
for all $f\in\cS_{\partial_-}(\R^3)$. In \cite{PullJMP} we have determined the
right-hand side of \eqref{equ:eq4} and thus we get finally
\begin{equation}\label{equ:e5}
	\LCD_m\tres_{x^+=0} = \frac{1}{4}(\delta(x_\bot)\otimes\epsilon(x^-)).
\end{equation}
\end{example}

The result \eqref{equ:e5} is related to what is called ``quantization on the
light cone'' \cite{LC-commutator, BrosPins}. Recall that a real scalar free field
$\phi=\phi_m$ obeys the commutator relation $[\phi(x),\phi(y)]=-iD_m(x-y)$
from which one obtains the canonical equal time commutator relations
$$
	[\phi(0,\bx),\phi(0,\by)]=0,\qquad
	[\phi(0,\bx),\pi(0,\by)]=i\delta(\bx-\by)
$$
by restricting $D_m$ and $\partial_0D_m$ to $\{x^0=0\}$; as usual we have set
$\pi(x)=\partial_0\phi(x)$. In light cone quantum field theory the field
$\phi$ is quantized by demanding commutator relations for fixed LC-time
$x^+=y^+=0$, where $x^+=\frac{1}{\sqrt 2}(x^0+x^3)$ and
$y^+=\frac{1}{\sqrt 2}(y^0+y^3)$, respectively. Transforming the covariant
field $\phi(x)$ to light cone variables
$\LCphi(\LCx)=(\phi\circ\kappa^{-1})(\LCx)$, the commutator relation
of light cone quantum field theory reads \cite{LC-commutator}
$$
	[\LCphi(0,\LCbx),\LCphi(0,\LCby)]=\frac{1}{4i}\left(
	\delta(x_\bot-y_\bot)\otimes\epsilon(x^- - y^-)\right).
$$
Furthermore, in light cone physics it is generally assumed that
the canonical field $\LCpi=\partial_+\LCphi$ is superfluous and need not
be considered through quantization. This is in accordance with our
transformation laws \eqref{equ:e1} and \eqref{equ:e2}, \eqref{equ:e3}
where the tame restriction $\LCu\tres_{x^+=0}$ of a solution $\LCu\in\cS'(\R^{1+n})$ of 
the LCKG-equation determines both $u|_{x^0=0}$
and $\partial_0u|_{x^0=0}$, and vice verse, where $u=\LCu\circ\kappa$.
Now since $[\phi(x),\phi(y)]=-iD_m(x-y)$ we obtain
$[\LCphi(\LCx),\LCphi(y)]=-i\LCD_m(\LCx-\LCy)$. Since $\LCD_m(\LCx)$ has
no canonical restriction to $\{x^+=0\}$ it is not surprising that one gets
into trouble in trying to restrict canonically $\LCphi(\LCx)$ to $\{x^+=0\}$
-- this has been an crucial problem in light cone physics \cite{PullJMP}.
If we consider the ``tame'' restriction \cite{PullJMP} $\LCphi(0,\LCbx)$ of the covariant 
LC-field $\LCphi(\LCx)$ then we obtain the canonical equal LC-time
commutator relation as $-i$ times the tame restriction
of $\LCD_m$ at $\LCbx-\LCby$ which, according to \eqref{equ:e5}, equals
$\frac{-i}{4}(\delta(x_\bot-y_\bot)\otimes\epsilon(x^- - y^-))$.
%
\section{Conclusions and Outlook}
%
In this paper we have investigated the characteristic Cauchy problem
of the Klein-Gordon equation \eqref{eq:char_Cauchy} which was motivated
from light cone quantum field theory.
As yet it has been an open question whether one can force
uniqueness within the space of all (physical) solutions. We solved this problem
by introducing the function spaces $\cS_{p^+}(\R^n)$ and
$\cS_{\partial_-}(\R^n)$ which appear as subspaces of the Schwartz space
$\cS(\R^n)$ and which are related to each other by the Fourier transformation
on $\cS(\R^n)$. To see that our consideration also includes the physical
case, recall (cf.\ \cite{Schweber}) that the ``physical'' solutions which correspond 
to the positive energy one-particle states of a real scalar free quantum field
are of the form
\begin{equation}\label{eq:Fock1}
	\phi(x^0,\bx)=\frac{1}{(2\pi)^3}\int\frac{d^3\bp}{2\omega(\bp)}
	\left( a(\bp) e^{-i(x^0\omega(\bp)-\bx\cdot\bp)} +
		a^+(\bp) e^{i(x^0\omega(\bp)-\bx\cdot\bp)}
	\right)
\end{equation}
where $a(\bp),a^+(\bp)\in\cL^2(\R^3,d^3\bp)$; notice that $\cS(\R^3)$ is
a dense subspace of $\cL^2(\R^3,d^3\bp)$. It follows from \eqref{eq:Fock1}
that the restrictions
$\phi|_{x^0=0}$ and $\partial_0\phi|_{x^0=0}$ are in $\cL^2(\R^3,d^3\bp)$
and, by the general result, $\phi$ is uniquely determined by these
restrictions. Going over to light cone
coordinates these solutions correspond to solutions (of the LCKG-equation)
of the form
\begin{eqnarray}\label{eq:Fock2}
	\LCphi(x^+,\LCbx) & = & (\phi\circ\kappa^{-1})(x^+,\LCbx)=
	\nonumber \\
	& = & \frac{1}{(2\pi)^3}\int_{p^+>0}\frac{d^2\LCbp}{2p^+}\left(
	b(\LCbp)e^{-i(x^+\LComega(\LCbp)+x^-p^+-x_\bot\cdot p_\bot)} +
	b^+(\LCbp)e^{i(x^+\LComega(\LCbp)+x^-p^+-x_\bot\cdot p_\bot)}
	\right)\\
	& = & \frac{1}{(2\pi)^3}\int\frac{d^3\LCbp}{2|p^+|}
	\LCb(\LCbp)e^{-i(x^+\LComega(\LCbp)+x^-p^+-x_\bot\cdot p_\bot)}
	\nonumber
\end{eqnarray}
where
$b(\LCbp),b^+(\LCbp)$ are in $\cL^2(\R^3,\frac{\Theta(p^+)}{2|p^+|}d^3\LCbp)$
and $\LCb(\LCbp)=\Theta(p^+)b(\LCbp)+\Theta(-p^+)b^+(-\LCbp)$ is in
$\cL^2(\R^3,d^3\LCbp/2|p^+|)$.
This follows immediately from our transformation law in 
Proposition~\ref{prop:transformation2} if one takes into account
that $\nu^*_{>0}\cL^2(\R^3,d^3\bp)=
\cL^2(\R^3,\frac{\Theta(p^+)}{2|p^+|}d^3\LCbp)$; notice that
$\cS_{p^+>0}(\R^3)=\nu^*_{>0}\cS(\R^3)$ is a dense subspace of
$\cL^2(\R^3,\frac{\Theta(p^+)}{2|p^+|}d^3\LCbp)$. We obtain from
\eqref{eq:Fock2} that the tame restriction $\LCphi\tres_{x^+=0}$ is the
Fourier transform of a function in $\cL^2(\R^3,\frac{1}{2|p^+|}d^3\LCbp)$.
Our general result says that this tame restriction is in
$\cS'_{\partial_-}(\R^3)$ and determines $\LCphi$ uniquely.

It is also possible to generalize the definitions of the spaces
$\cS_{p^+}(\R^n)$ and $\cS_{\partial_-}(\R^n)$ in the following way:
Recall that $\cS_{p^+}(\R^n)$ is the (complex) vector space of all 
$C^\infty$-functions such that
\begin{equation}\label{eq:c1}
	||f||_{Q,\alpha}=\sup_{\LCbp\in\R^n\setminus\{p^+=0\}}
		|Q(\LCbp)\partial^\alpha(\LCbp)|<\infty
		\qquad Q\in\C[p^+,p_\bot]_{p^+},~\text{$\alpha$ multi-index,}
\end{equation}
topologized by the seminorms in the left-hand side of\eqref{eq:c1};
$\cS_{p^+}(\R^n)$ is a subspace of $\cS(\R^n)$ and the topology on
$\cS_{p^+}(\R^n)$ coincides with the subspace topology induced by $\cS(\R^n)$. At first
sight the definition of $\cS_{p^+}(\R^n)$ looks
artificial. However, the following consideration shows that
the definition is as natural as that of $\cS(\R^n)$. Let
$F(x)\in\C[X]$, $X=(X_1,\ldots,X_n)$, be a complex polynomial. Consider
the (affine) variety $V(F)=\{x\in\C^n:F(x)=0\}$ which is a hypersurface
of the affine space $\A^n$. The complement $D(F)=\A^n\setminus V(F)$ is
called a distinguished open subset of $\A^n$; $D(F)$ is itself a variety.
The ring of regular functions on $D(F)$ is $\C[X]_F$ which consists of
formal fractions $q/F^r$, $q\in\C[X]$, $r\geq 0$. Any element of
$q/F^r\in\C[X]_F$ induces canonically a (complex-valued) function $x\mapsto q(x)/F^r(x)$
on $D(F)$. Now let $\cS_F(\R^n)$ be the vector space of all
$C^\infty$-functions on $D_\R(F)=D(F)\cap\R^n$ such that
\begin{equation}\label{eq:c2}
	||f||^F_{Q,\alpha}=\sup_{x\in D_\R(F)}
		|Q(x)\partial^\alpha f(x)|<\infty
		\qquad Q\in\C[X]_F,~\text{$\alpha$ multi-index,}
\end{equation}
topologized by the seminorms in the left-hand side of \eqref{eq:c2}; we
call the dual space $\cS'_F(\R^n)$ the space of {\em $F$-tempered
distributions} \cite{Pull_future2}. Notice that any element of $\cS_F(\R^n)$ can be extended
(by zero) to a $C^\infty$-function on $\R^n$.
If $F=\text{const.}$\ is a constant polynomial then we recover the
Schwartz space $\cS(\R^n)$, i.e., $\cS_F(\R^n)=\cS(\R^n)$. Hence $\cS_F(\R^n)$
is a natural generalization of the Schwartz space $\cS(\R^n)$. In the
solution of the characteristic Cauchy problem of the Klein-Gordon equation
we have made use of $\cS_F(\R^n)$ where $F(\LCbp)=p^+$. If $F_1,F_2\in\C[X]$
are polynomials such that $D(F_1)\subset D(F_2)$, then canonically
$\C[X]_{F_2}\subset\C[X]_{F_1}$ and one obtains
$\cS_{F_1}(\R^n)\subset\cS_{F_2}(\R^n)$. Hence, if especially
$F_2=\text{const.}$\ is a constant polynomial then we obtain
$\cS_F(\R^n)\subset\cS(\R^n)$ for any polynomial $F\in\C[X]$. Thus one
can consider the preimage of $\cS_F(\R^n)$ in $\cS(\R^n)$
under the Fourier transformation; this preimage is denoted $\cS_{F(\partial)}(\R^n)$. 
Furthermore, the dual spaces $\cS'_F(\R^n)$ and
$\cS'_{F(\partial)}(\R^n)$ depend functorially on $D(F)$, i.e.,
they are presheaves. These spaces may be helpful in constructing non-canonical 
restrictions (and products) of distributions and in studying the characteristic
Cauchy problem of partial differential equations in general.
\vspace{.5cm}\\
\noindent
\textit{Acknowledgement.}  I would like to thank Ernst Werner for helpful
discussions and comments.


\end{document}